\documentclass[twocolumn,showpacs,preprintnumbers,amsmath,amssymb]{revtex4}

\usepackage{graphicx}
\usepackage{amssymb}
\usepackage{dcolumn}
\usepackage{bm}
\usepackage{epstopdf}

\begin{document}
%
%

\title{Haldane Sashes in Quantum Hall Spectra}

\author{A.H. MacDonald}
\affiliation{Department of Physics, The University of Texas at Austin, Austin, Texas 78712, USA}

\begin{abstract}
We show that the low-temperature sash features in the lowest Landau-level (LLL) tunneling density-of-states (TDOS) 
recently discovered by Dial and Ashoori are intimately related to the discrete 
Haldane-pseudopotential interaction energy scales that govern fractional quantum Hall physics.  
Our analysis is based on expressions for the tunneling density-of-states which become exact 
at filling factors close to $\nu=0$ and $\nu=1$, where the sash structure is most prominent.  
We comment on other aspects of LLL correlation physics that can be revealed by accurate 
temperature-dependent tunneling data. 
\end{abstract}


\maketitle

\noindent
{\em Introduction}---Many-electron correlation physics within a partially filled Landau level has been an enduring
source of new physics for several decades.  In recent years the even-denominator incompressible states\cite{nu52} 
that occur when the $n=1$ Landau level is partially filled, and the smectic states\cite{smectic} that occur for 
$n > 1$ have received particular attention.  Correlation physics in the quantum Hall regime has most 
frequently been probed experimentally by studying transport properties of very high quality
GaAs/(Al,Ga)As two-dimensional electron gases (2DEGs) and identifying the filling factors at which 
the quantum Hall effect occurs.  The quantum Hall effect is a transport anomaly characterized by 
plateaus in the Hall conductance and vanishing longitudinal resistance over a finite range 
of magnetic field strength or electron density.  Its occurrence signals\cite{leshouches} a jump in the chemical 
potential at a density which depends on magnetic field, and localization of the gapped charged excitations of the 
corresponding incompressible state.  

Many of the correlated electron states that 
occur in quantum Hall systems do not have parallels elsewhere in physics and represent qualitatively 
new classes of electron behavior.  Some, including notably the $\nu=1/2$ composite-fermion liquid state\cite{composite_fermion}
state, are not characterized by charge gaps.  Others have gaps, but also have poorly understood properties unrelated to their low-energy
charged excitations.  In both cases transport is an unsatisfying probe of the properties of interest.  
Unfortunately experiments that are complementary to transport 
have often not been readily available, mostly because of experimental challenges associated 
with the buried 2DEG location.  
The present article in motivated by the $n=0$ tunneling spectra observations recently reported 
by Dial and Ashoori\cite{Dial_Nature} which have revealed, in addition to the Coulomb gap behavior already evident 
in earlier data\cite{Eisenstein,Ashoori_Old}, a pattern of sharp structure at 
high energies that is most pronounced for filling factors near $\nu=0$ and $\nu=1$.  The authors refer to these 
unanticipated structures as {\em sashes}.  In this Letter we show that they occur because
correlation physics in the quantum Hall regime is governed by a small set of 
discrete energy scales ($V_{m}$) known as Haldane pseudopotentials\cite{haldane,leshouches}.

\noindent
{\em TDOS at small Landau level filling factor}---We start by presenting some exact results 
for the small $\nu$, $T=0$, limit of the  LLL TDOS ($A(\epsilon)$) of 
a disorder-free 2DEG.  The TDOS is the sum of electron-addition and electron-removal 
contributions given at $T=0$ by\cite{mahan,caveatsymmetricgauge}  
\begin{equation} 
A_{+}(\epsilon) = \sum_{n} |\langle \Psi_n(N+1) | c_{m}^{\dagger} | \Psi_0(N)\rangle|^2 \; \delta(\epsilon - E_{n,0}),
\end{equation} 
and 
\begin{equation} 
A_{-}(\epsilon) = \sum_{n} |\langle \Psi_n(N-1) | c_{m} | \Psi_0(N)\rangle|^2 \; \delta(\epsilon + E_{n,0}).
\end{equation} 
Here $|\Psi_n(N \pm 1)\rangle$ are exact eigenstates of the $N \pm 1$ particle system,
$| \Psi_0(N)\rangle$ is the ground state of the $N$-particle system, and $E_{n,0} = E_n(N \pm 1)- E_0(N)$ is the 
energy of a charged excitation.   In the absence of disorder, it follows from translational invariance\cite{caveatsymmetricgauge,haussmann} that $A(\epsilon)$ 
is independent of the single-particle angular momentum label $m$ and depends only on the filling factor $\nu=N/N_{LL}$.  ($N_{LL}$ is the 
number of states per Landau level in a finite area system.)   When energies are measured from the
chemical potential, the differential conductivity\cite{mahan} at bias voltage $V$ between a LLL 2DEG 
and a counter-electrode that is separated by a thin tunnel barrier and has a smooth density-of states is equal to $A(eV)$ up to a constant factor.  
When an electron is suddenly added to or removed from the ground state, 
correlations are disturbed and the resulting state is not an eigenstate of the 
many-electron Hamiltonian.  Tunneling spectroscopy experiments measure the energy probability distribution  
in the sudden state.  Note that $A_{+}(\epsilon)$ is non-zero only for $\epsilon > \mu_{N} = E_0(N+1)-E_0(N)$, whereas 
$A_{-}(\epsilon)$ is non-zero only for $\epsilon < \mu(N-1) = E_0(N)-E_0(N-1) \le \mu(N)$.

We now derive analytic results for $A(\epsilon)$ that are valid when the LLL is 
nearly filled or nearly empty by exploiting particle-hole symmetry\cite{particlehole} and the very simple properties of $N=2$ LLL eigenstates.   
In the absence of disorder, the ground state of a $N=1$ LLL system is degenerate; electrons can  
occupy any angular momentum from $m=0$ to $m=N_{LL}-1$. 
We choose our zero of energy so that the kinetic  
energy in the LLL is $0$.  The $N=1$ ground state energy is then $-\Delta_z/2$ where 
$\Delta_{z}$ is the Zeeman splitting between majority and minority spin energies.
The electron removal part of the spectral function for $N=1$ is a delta-function at this 
energy with weight $N_{LL}^{-1}$.  Two-particle electron states are either the product of a triplet spin state and an orbital state that changes 
sign under particle interchange, or the product of a singlet spin state and an orbital state that is 
invariant under particle interchange.  Both types of two-particle orbital states are conveniently obtained by repeated 
application of center-of-mass (COM) and relative angular momentum raising operators\cite{leshouches}:
\begin{eqnarray} 
\label{eq:convertion}
b_{R}^{\dagger} &=& \frac{b_1^{\dagger} + b_2^{\dagger}}{\sqrt{2}} \nonumber \\ 
b_{r}^{\dagger} &=& \frac{b_1^{\dagger} + b_2^{\dagger}}{\sqrt{2}} 
\end{eqnarray} 
where $b_i^{\dagger}$ is the LLL angular momentum raising operator for particle $i$, 
$b_{R}^{\dagger}$ raises the center-of-mass angular momentum $M$ and $b_{r}^{\dagger}$ 
raises the relative angular momentum $k$.  The $N=2$ orbital states labeled by $M$ and $k$ are
\begin{equation} 
|M,k\rangle = \frac{(b_{R}^{\dagger})^M (b_{r}^{\dagger})^k}{\sqrt{M!k!}}  \; \;  |0,0\rangle .
\end{equation} 
Here $|0,0\rangle$ is the two-particle state in which both individual 
angular momenta, the COM angular momentum, and the relative angular momentum,are all equal to $0$. 
Since the interaction Hamiltonian acts only on the relative degree-of-freedom and is 
diagonal in relative angular momenta when the 2DEG is isotropic, these states are two-particle 
Hamiltonnian eigenstates with eigenenergy $V_{k} -\Delta_{z} S_z$.  Here
$S_z$ is the component of total spin along the field direction and $V_{k}$,
the expectation value of the pair interaction in relative-angular-momentum state $k$, 
is the $k$'th Haldane pseudopotential.
Correlation physics in the quantum Hall regime depends almost entirely 
on the relative values of the first few Haldane pseudopotentials.

We detail our electron addition spectral function calculation
only for the case in which the angular momentum of the added electron 
$m_{2}=0$; it is easy to verify that identical results are obtained for 
any value of $m_2$ as required by translational invariance.
Our spectral-function calculation proceeds by averaging  
over all possible values $m_1$ of the $N=1$ state angular momentum.  We consider 
first the case in which the spin of the added electron is parallel to the $N=1$ ground 
state spin. The anti-symmetrized two-particle state created upon electron addition, 
\begin{equation}
\label{eq:inst} 
|\Psi_{m_2,m_1}\rangle = \frac{(b_1^{\dagger})^{m_1} \, (b_2^{\dagger})^{m_2} - (b_1^{\dagger})^{m_2} \, (b_2^{\dagger})^{m_1}}{\sqrt{2 m_{1}! m_{2}!}} \; |0,0\rangle,
\end{equation} 
is not an eigenstate of the Hamiltonian.   In order to evaluate the spectral function we need to expand this state in terms of the two-particle eigenstates which have 
definite relative angular momentum $k$.  This is easily accomplished using the relationship between single-particle and COM-relative angular momentum
raising operators (Eq.~(\ref{eq:convertion})).  After a bit of algebra we find that the probability of obtaining a state with odd relative angular momentum $k$
upon adding an electron with $m_2=0$ to a single electron state with angular momentum $m_1$ is
\begin{equation} 
P_k(m_1) = \frac{2}{k!} \; \frac{m_{1}!}{2^{m_{1}} (m_{1}-k)!}.
\end{equation} 
(Even relative angular momenta do not appear in the eigenstate expansion of the parallel spin sudden state.)
For example, adding a $m_2=0$ electron to a $N=1$ state with $m_1=3$ yields a state with relative angular momentum $1$ with probability $P_1(3)=3/4$ 
and a state with relative angular momentum $3$ with probability $P_3(3)=1/4$.  
Noting that $\sum_{m_1} P_k(m_1) = 4$ for all $k$, we obtain for the following result for the parallel spin contribution to the $N=1$ electron addition 
spectral function:
\begin{equation} 
\label{eq:N1parallel}
A_{+}^{\parallel}(\epsilon) = \frac{4}{N_{LL}} \; \sum_{k \in {\rm odd}} \; \delta(\epsilon- V_{k} + \Delta_{z}/2), \; 
\end{equation} 
where the maximum value of $k$ is $N_{LL}/2$. 
 
When the added electron and the $N=1$ ground state electron have opposite spin, the symmetrized 
two-particle states have a spin-factor that is the $S_z=0$ member the two-particle singlet for   
even $k$ and the $S_z=0$ member of the two-particle triplet for odd $k$.  The normalization factor of
these spin-states leads to a result for $P_k(m_1)$ that is smaller than the parallel spin result 
by a factor of two.  The end result is that for opposite spin addition and $N=1$ 
\begin{equation} 
\label{eq:N1opposite}
A_{+}^{{\rm opp}} = \frac{2}{N_{LL}} \sum_{k} \delta(\epsilon- V_{k} - \Delta_{z}/2). \;  
\end{equation}
The TDOS consists of peaks located, apart from small Zeeman energy contributions, at both even and odd $k$ 
Haldane pseudopotential energies.  

For values of $\nu$ close to zero, the typical distance between
correlated ground state electrons is large.  We can therefore view a system
with area $A$ as consisting of a $N$ subsystems with area $A/N$ within which 
interactions with other electrons can be neglected.  The $N=1$ results 
for spectral functions contributions near energy $V_k$ therefore 
apply at small but finite $\nu$ if we replace $N_{LL}^{-1}$ in Eqs.~(\ref{eq:N1parallel}) and (\ref{eq:N1opposite}) by  
$(N_{LL}/N)^{-1}=\nu$, provided that $k $ is small compared to $N_{LL}/2N = 1/2\nu$. 
Added electrons will have a significant probability of 
forming high-energy small-$k$ relative angular momentum 
states only if they are added at a position close to 
that of a ground state electron, and this becomes more likely as
$\nu$ increases.  It follows that a peak in the spectral function 
with weight proportional to $\nu$ is expected near $\epsilon = V_{k}$ over a range of $\nu$ that decreases with $k$.

In the tunneling data of Dial and Ashoori\cite{Dial_Nature} only the $k=0$ and $k=1$ peaks are apparent.
Higher $k$ peaks that would be expected to be visible over narrower ranges 
of $\nu$ are evidently obscured by disorder, which is of course present even in these very 
high quality samples.   We refer to these peaks in the TDOS as {\em Haldane sashes}. 
From Ref.[\onlinecite{Dial_Nature}] we 
can read off experimental values for the $k=0$ and $k=1$ Haldane pseudospotentials:  $V_{0} \approx 9 meV$, and 
$V_{1} \approx 6 meV$. The $V_{1}$ peak feature appears to be 
approximately three times stronger than the higher energy $V_{0}$ peak feature as predicted by this 
theoretical analysis.     

The Haldane sashes should broaden as $\nu$ increases.
They are nevertheless evident over a fairly broad range
of $\nu$ and provide valuable qualitative insight into LLL electronic correlations.  
The $k=1$ sash feature, for example, is evident up to $\nu=1/3$ where its lower 
edge approaches the Fermi level.  This behavior is expected since the possibility of making  
charged excitations that avoid relative-angular-momentum $k=1$ is absent\cite{laughlin,leshouches}for $\nu > 1/3$.
Similarly the $k=0$ feature, associated with adding an opposite spin electron at the same position as an existing ground
state electron, is evident for all $\nu \le 1$.  At $\nu=1$ its lower edge approaches the 
Fermi energy.  The $k=1$ sash is not visible over the range  
$1/3 < \nu < 2/3$ where  
composite-fermion physics reigns, demonstrating that the relationship
between spectra and spatial correlations is more subtle in the composite fermion regime. 
Although composite fermion physics is not revealed by the $T=0$ TDOS,
we anticipate that it will emerge in its T-dependence as explained below.     

Provided that the ground state is maximally 
spin-polarized at all filling factors\cite{spinpolrefs}, 
particle-hole symmetry\cite{haussmann,particlehole} 
implies that the parallel spin contribution to $A_{+}(\epsilon)$ at filling factor $\nu$ 
equals $A_{-}(-\epsilon)$ at filling factor $1-\nu$.
It follows that hole-pair Haldane sashes 
are present in the electron removal part of the 
spectral function near $\nu=1$.  Hole-pair Haldane sashes
are also evident in Ref.[\onlinecite{Dial_Nature}].   

\noindent
{\em TDOS and Correlation Energies}---The progress reported in Ref.[\onlinecite{Dial_Nature}] 
motivates a reexamination of the relationship between the TDOS and correlation energies discussed 
previously\cite{haussmann} by Haussmann {\em et al.}.  This relationship is simplest when only one spin-component is involved and,
for this reason, we now focus on the electron-removal TDOS for $\nu < 1$. Recognizing that 
energies are in general known only relative to the chemical potential, we rewrite the energy expression derived 
by Haussmann {\em et al.} in the form
\begin{equation} 
\label{eq:sumrule}
\int_{-\infty}^{\infty} d \xi \, \xi \, A_{-}(\xi) n_F(\xi) \equiv S_{-}(\nu) =  2 \epsilon_0(\nu) - \mu (\nu),
\end{equation} 
where $\xi$ is energy measured from the chemical potential, $n_F(\xi)=\theta(-\xi)$ is the $T=0$ Fermi factor,
$\epsilon_0(\nu)$ is the ground state energy per LLL state relative to the LLL single-particle energy and 
$\mu = \epsilon_{0}'(\nu)$
is the chemical potential.  Note that in evaluating $A_{-}(\xi)$ from experimental tunneling data, uncertainty related to 
tunneling matrix-element factors can be mitigated by applying the sum rule 
\begin{equation}
\int_{-\infty}^{0} d \xi \, A_{-}(\xi) n_F(\xi) = \nu. 
\end{equation}  
Eq.(~\ref{eq:sumrule}) can be rewritten in the form, 
\begin{equation} 
\frac{d}{d \nu} \; \frac{\epsilon_0(\nu)}{\nu^2} = - \frac{S_{-}(\nu)}{\nu^3},
\end{equation}
and integrated over filling factor to extract the filling factor dependence of the ground state 
energy and chemical potential directly from the tunneling data.  For very small $\nu$
we expect the ground state energy to be given accurately by the classical triangular lattice
value\cite{maradudin} 
\begin{equation} 
\epsilon_0(\nu) \to \epsilon_{WC}(\nu) = -0.782133 \,  \frac{e^2}{\epsilon \ell}  \, \nu^{3/2}.
\end{equation} 
where $\ell$ is the magnetic length.
It follows that in the same limit $S_{-}(\nu) \propto \nu^{3/2}$ and that integrals of  
$S_{-}(\nu)/\nu^3$ do not converge when the lower limit extends to $\nu=0$. 
(The $\nu^{3/2}$ small $\nu$ behavior of $S_{-}(\nu)$ is evident in Ref.[\onlinecite{Dial_Nature}.])   
Ground state energies must therefore be determined either by connecting to the classical 
triangular lattice value at a small filling factor $\nu^*$ using,
\begin{equation} 
\epsilon_0(\nu) = \nu^2 \Big[\; \frac{\epsilon_{WC}(\nu*)}{(\nu^*)^{2}} - \int_{\nu^*}^{\nu} \,d \nu' \; \frac{S_{-}(\nu')}{\nu'^{3}} \; \Big], 
\end{equation} 
or by relating the correlated fractional filling factor ground state energy to the 
full Landau level ground state energy 
using
\begin{equation} 
\epsilon_0(\nu) = \nu^2 \Big[\; \epsilon_0(\nu=1) +  \int_{\nu}^{1} \,
d \nu' \; \frac{S_{-}(\nu')}{\nu'^{3}}\;  \Big].
\end{equation} 
Once $\epsilon_0(\nu)$ is known, $\mu(\nu)$ can be obtained using Eq.(~\ref{eq:sumrule}).  

\noindent
{\em Discussion}---Correlation physics in a Landau level is known to be highly sensitive to 
Haldane pseudopotential $V_k$ values; the qualitative differences between $n=0$ composite-fermion, 
$n=1$ non-abelian quasiparticle, and $n \ge 2$ density-wave physics are 
associated with relatively modest changes in $V_k$ ratios.  We have 
shown that tunneling spectroscopy can be used to measure the most important $V_k$ values.
Since the numerical values of 
these parameters can be difficult to estimate accurately on the basis of 
theoretical considerations because of uncertainties related to quantum 
well width and quantum fluctuations involving higher Landau levels and higher subbands, the 
ability to measure their values using tunneling spectroscopy is a valuable advance.  The numerical 
values measured by Dial and Ashoori, $V_0 \approx 9 {\rm meV}$ and 
$V_1 \approx 6 {\rm meV}$, are in the expected range.     

One important experimental finding of Dial and Ashoori is that Haldane sashes 
survive only up to temperatures that are quite small compared to the energies at which they appear.  
This behavior is expected since, as we have explained, the sharpness of the Haldane sashes is dependent on correlations which 
keep the electrons well separated.  The small filling factor Wigner cyrstal state, for example, 
melts\cite{wcmelting} when $k_{B} T \gtrsim 0.005 (e^2/\epsilon \ell) \nu^{1/2}$.  
Sash features in the TDOS reflect strong 
correlations which develop only as the crystallization temperature is approached.

The temperature dependence of the TDOS can also shed light\cite{caveat} on
thermodynamic properties that 
are normally not accessible in 2DES's and could lead to valuable new  
insights into LLL correlation physics. 
Since Eq.(~\ref{eq:sumrule}) remains valid\cite{mahan} at finite $T$,   
tunneling data can  be used to determine $\epsilon_0(\nu)$ and $\mu(\nu)$ as a function of T.
The temperature dependence of energy yields the heat capacity,
\begin{equation}  
C = N_{LL} \frac{\partial \epsilon_0}{\partial T} = T \frac{\partial S}{\partial T}, 
\end{equation} 
and therefore the temperature-dependence of the entropy.  Unlike the TDOS, the 
heat capacity and the entropy are determined by the excitation spectrum of the 
system alone, and are not influenced by the overlap\cite{coulombgap} between 
charged excitations and bare electron excitations.  
According to composite fermion theory, the 
heat capacity at Landau level filling factor $\nu=1/2$ is given by    
\begin{equation} 
\label{eq:heatcapacity} 
\frac{C}{N_{LL}} = \frac{\pi^2 k_B^2 T}{3} \; \frac{ m^*_{cf} \ell^2}{\hbar^2} 
\end{equation} 
where $m^*_{cf}$ is the composite fermion effective mass.  
Verification of Eq.(~\ref{eq:heatcapacity}) could directly establish the composite fermion ansatz.
The property that the 
heat capacity depends on $m^*_{cf}$ alone, independent of complex quasiparticle 
normalization factors, is analogous to the corresponding Fermi liquid property.  

The temperature-dependence of the chemical potential could also be 
very revealing, especially for incompressible states like the one thought 
to be responsible for the quantum Hall effect at 
filling factor $\nu=5/2$ ($\nu=1/2$ in the $n=1$ Landau level), that have non-Abelian quasiparticles.
The theoretically expected quasiparticle statistics imply\cite{CooperStern,YangHalperin} finite entropy, 
\begin{equation} 
S = \frac{k_B |2 N - N_{LL}| \ln{2}}{2},  
\end{equation} 
above exponentially small temperatures at fractional filling factors near $\nu=1/2$ within a $n=1$ level.  
The dependence of entropy on particle number, which changes sign when
$\nu$ crosses $1/2$, can be 
obtained directly from chemical potential data using the thermodynamic identity
\begin{equation} 
\frac{\partial S}{\partial N} = \frac{\partial \mu}{\partial T}.
\end{equation} 

\noindent
{\em Acknowledgements}--- AHM thanks Oliver Dial and Ray Ashoori for a stimulating conversation
during EP2DS18.  This work was supported by the Welch Foundation and by the National Science Foundation under 
grant DMR-0606489.


\begin{references}
\bibitem{caveatsymmetricgauge} We couch our discussion in terms of symmetric gauge LLL states in 
which single-particle states have non-negative integer angular momentum labels.
\bibitem{caveat} 
The differential TDOS at finite temperature is proportional to 
$A(\xi)$ averaged over energy with a $\partial n_F(\xi)/\partial \xi$ weighting factor.  
It is in principle possible if necessary to correct for the thermal broadening and extract the underlying 
temperature-dependent spectral function.
\end{references}
\end{document}